\documentclass[aps, prd, twocolumn, nofootinbib, longbibliography, superscriptaddress]{revtex4-1}

\usepackage{epsfig}
\usepackage[top=30pt,bottom=30pt,left=48pt,right=46pt]{geometry}
\usepackage{enumitem}
\usepackage{amsmath}
\usepackage{amssymb}
\usepackage{graphicx}
\usepackage{booktabs}
\usepackage{epstopdf}
\usepackage{gensymb}
\usepackage{aas_macros}
\usepackage{multirow}
\usepackage{lipsum}

\usepackage[colorlinks=true,citecolor=blue,linkcolor=blue]{hyperref}
\usepackage[capitalise,noabbrev]{cleveref}

\usepackage{tikz,xcolor,hyperref}

\definecolor{lime}{HTML}{A6CE39}
\DeclareRobustCommand{\orcidicon}{\hspace{-4pt}
	\begin{tikzpicture}
		\draw[lime, fill=lime] (0,0) 
		circle [radius=0.16] 
		node[white] {\hspace{0.1mm}{\fontfamily{qag}\selectfont \tiny ID}};
		\draw[white, fill=white] (-0.07,0.1) 
		circle [radius=0.01];
	\end{tikzpicture}
	\hspace{-3.2mm}
}

\foreach \x in {A, ..., Z}{\expandafter\xdef\csname 
	orcid\x\endcsname{\noexpand\href{https://orcid.org/\csname orcidauthor\x\endcsname}
		{\noexpand\orcidicon}}
}

\begin{document}
	
	\title{First search for high-energy neutrino emission from galaxy mergers}
	
	\author{Subhadip Bouri\orcidA{}}
	\email{subhadipb@iisc.ac.in}
	\affiliation{Department of Physics, Indian Institute of Science, C.\,V.\,Raman Avenue, Bengaluru 560012, India}
	
	\author{Priyank Parashari\orcidB{}}
	\email{ppriyank@iisc.ac.in}
	\affiliation{Centre for High Energy Physics, Indian Institute of Science, C.\,V.\,Raman Avenue, Bengaluru 560012, India}
	
	\author{Mousumi Das\orcidD{}}
	\email{mousumi@iiap.res.in}
	\affiliation{Indian Institute of Astrophysics, Koramangala, Bengaluru 560034, India}
	
	\author{Ranjan Laha\orcidC{}}
	\email{ranjanlaha@iisc.ac.in}
	\affiliation{Centre for High Energy Physics, Indian Institute of Science, C.\,V.\,Raman Avenue, Bengaluru 560012, India}

	\date{\today}
	
	\begin{abstract}
 The exact sources of high-energy neutrinos detected by the IceCube neutrino observatory still remain a mystery. For the first time, this work explores the hypothesis that galaxy mergers may serve as sources for these high-energy neutrinos. Galaxy mergers can host very high-energy hadronic and photohadronic processes, which may produce very high-energy neutrinos. We perform an unbinned maximum-likelihood-ratio analysis utilizing the galaxy merger data from six catalogs and 10 years of public IceCube muon-track data to quantify any correlation between these mergers and neutrino events. First, we perform the single source search analysis, which reveals that none of the considered galaxy mergers exhibit a statistically significant correlation with high-energy neutrino events detected by IceCube. Furthermore, we conduct a stacking analysis with three different weighting schemes to understand if these galaxy mergers can contribute significantly to the diffuse flux of high-energy astrophysical neutrinos detected by IceCube. We find that upper limits (at $95\%$ CL) of the all flavor high-energy neutrino flux, associated with galaxy mergers considered in this study, at $100$ TeV with spectral index $\Gamma=-2$ are $1.11\times 10^{-18}$, $3.69 \times 10^{-19}$ and $1.02  \times 10^{-18}$ $\rm GeV^{-1}\,cm^{-2}\,s^{-1}\,sr^{-1}$ for the three weighting schemes. This work shows that these selected galaxy mergers do not contribute significantly to the IceCube detected high energy neutrino flux. We hope that in the near future with more data, the search for neutrinos from galaxy mergers can either discover their neutrino production or impose more stringent constraints on the production mechanism of high-energy neutrinos within galaxy mergers.
	\end{abstract}
	
	\maketitle

\section{Introduction}
Neutrinos{\footnote{In this work, we do not distinguish between neutrino and antineutrino}} are mysterious particles as they are ubiquitous in the Universe, yet they are very difficult to detect due to their feeble interaction with matter. Their elusive nature also makes them a great tool to probe various extreme environments in our Universe. Moreover, neutrinos exhibit a wide energy spectrum ranging from a few meV to $10^{20}$ eV, depending on the formation process of these neutrinos~\cite{Vitagliano:2019yzm}. 

Over the last decade, IceCube, a neutrino telescope located at the South Pole, has detected several high-energy astrophysical neutrino events in the $\mathcal{O}$(10 TeV) -- $\mathcal{O}$ (PeV) energy range~\cite{IceCube:2013cdw,Aartsen:2013jdh,IceCube:2014stg, Aartsen:2015knd,IceCube:2015qii,Stettner:2019tok, Stachurska:2019wfb, Aartsen:2020aqd,IceCube:2021uhz, Abbasi:2020jmh,IceCube:2021rpz,Abbasi:2024jro,IceCube:2024pov,IceCube:2025zyb}. The number of astrophysical neutrinos detected by IceCube includes $127\pm12$ electron neutrinos, $22\pm2$ muon neutrinos, and $80\pm7$ tau neutrinos from 2010–2015 cascade data \cite{Aartsen:2020aqd}. Additionally, 60 events were detected above 60 TeV in the 7.5-year High Energy Starting Events (HESE) sample, among which 48.4 are expected to be of astrophysical origin\,\cite{Abbasi:2020jmh}, and 680 astrophysical muon neutrinos were detected over 10.3 years of track data\,\cite{Abbasi:2024jro}. Recently, ANTARES and Baikal-GVD collaborations have also reported the observation of diffuse high energy astrophysical neutrino flux~\cite{ANTARES:2017srd,Baikal-GVD:2022fis}. The detection of diffuse high-energy astrophysical neutrinos has emerged as a groundbreaking milestone in the field of astroparticle physics\,\cite{Ahlers:2018fkn,Klein:2018fnn,Troitsky:2021nvu,Halzen:2022pez,Winter:2024zpk}. This discovery has provided us with an innovative approach to investigate the Universe and a novel tool for probing new physics. A significant amount of work has been done to search for sources of these high-energy neutrinos. So far, only two sources, TXS 0506+056~\cite{IceCube:2018cha, IceCube:2018dnn} and NGC 1068~\cite{Aartsen:2019fau,IceCube:2022der}, have been confirmed by the IceCube collaboration to be sources of such high-energy neutrinos. However, both these sources can only explain 1$\%$ of the total diffuse flux of high-energy astrophysical neutrinos~\cite{IceCube:2018cha,IceCube:2022der}. Moreover, several studies were carried out to search for the sources of these high-energy neutrinos from various known astrophysical objects like TXS 0506+056 and NGC 1068. However, these analyses found that no more than 20$\%$ of high-energy astrophysical neutrino flux can arise from such sources~\cite{Aartsen:2016lir, Hooper:2018wyk, Smith:2020oac}. Few studies have also found that tidal disruption events~\cite{Stein:2020xhk}, Cygnus region of Milky Way~\cite{Neronov:2023hzu}, and NGC 7469~\cite{Sommani:2024sbp} are associated with observed high-energy neutrinos. Recently, IceCube collaboration has also reported the observation of high energy neutrinos from the Milky Way galactic plane~\cite{IceCube:2023ame}. Other sources that have been searched as neutrino emitters are blazars~\cite{Aartsen:2016lir, Hooper:2018wyk, Yuan:2019ucv, Luo:2020dxa, Smith:2020oac,Bellenghi:2023yza,IceCube:2022zbd, IceCube:2023htm,Plavin:2022oyy,2020arXiv200201661M,2022A&A...658L...6D,Plavin:2020mkf,ANTARES:2023lck}, gamma-ray bursts~\cite{Waxman:1997ti, Abbasi:2009ig, Abbasi:2011qc, Abbasi:2012zw, Aartsen:2014aqy, Aartsen:2016qcr, Aartsen:2017wea,IceCube:2022rlk,IceCube:2023woj}, fast radio bursts~\cite{IceCube:2017fpg,2017ApJ...845...14F,Kheirandish:2019dii,IceCube:2019acm,Desai:2021dpm,IceCube:2022mjy}, novae~\cite{IceCube:2022lnv}, active galactic nuclei~\cite{Zhou:2021rhl,IceCube:2021pgw,Halzen:2023usr,McDonough:2023ngk}, choked jet supernovae~\cite{Senno:2017vtd, Esmaili:2018wnv,Chang:2022hqj,Tagawa:2023hli,IceCube:2023esf}, pulsar wind nebulae~\cite{Aartsen:2020eof}, star forming galaxies~\cite{Chang:2014sua,Emig:2015dma,Bechtol:2015uqb,2016JCAP...12..021M,2016arXiv160703361C}, ultraluminous infrared galaxies~\cite{IceCube:2021waz}, seyfert galaxies~\cite{Neronov:2023aks},  galactic X-ray binaries~\cite{IceCube:2022jpz}, radio pulsars~\cite{Pasumarti:2022yba,Pasumarti:2023apw}, giant molecular clouds~\cite{2024arXiv240105863R}, galaxy clusters~\cite{IceCube:2022ccm}, and dwarf spheroidal galaxies~\cite{Guo:2023axz}. Moreover, there have been various theretical works that propose numerous candidates as sources of high-energy neutrinos~\cite{Murase:2006mm,Kashiyama:2013qet,Murase:2017pfe,Murase:2018okz,Yuan:2020oqg,Xiao:2016rvd,Senno:2015tra,Yoshida:2022idr,Fang:2020bkm,Reusch:2021ztx,Ambrosone:2021aaw,Kuze:2022uzz,Chang:2022bro,Sarmah:2022vra,Guarini:2022uyp,Jaroschewski:2022gdy,Kalashev:2022scs,Zhou:2022jle,Sridhar:2022uis,Buson:2023irp,Pitik:2023vcg,Sarmah:2023sds,Troitsky:2023nli,Ambrosone:2023kqa,Ray:2023awr,Sarmah:2023xrm}. Additionally, there have also been some studies which investigate the correlation of the observed high-energy neutrinos with gamma-rays and gravitational waves~\cite{2018arXiv180109545T,2019GCN.25775....1I,2019ICRC...36..932K,2019ICRC...36..787S,Fermi-LAT:2019hte,VERITAS:2021mjg,IceCube:2021jwt,PhysRevD.103.103020,2022PhRvD.106h3024L,IceCube:2022heu,Li:2022vsb,Negro:2023kwv,ANTARES:2018bmu,IceCube:2022mma, IceCube:2021ddq,Kumar:2023yaz,Klinger:2023zzv, Garrappa:2024zsm,2024A&A...681A.119R,Gagliardini:2024een,Li:2024gnb}. Despite significant efforts to understand the origin of high-energy neutrinos, their exact sources remain mysterious, and therefore, they continue to be a subject of further investigation.

In this study, we explore the possibility of galaxy mergers being the potential source of high-energy astrophysical neutrino flux observed by IceCube. It has been shown theoretically that galaxy mergers can produce ultrahigh energy neutrinos through shock mechanism~\cite{1993ICRC....2..341C,Jones:1997qr,2010A&A...524A..27L,2014ApJ...790L..14K,2018ApJ...857...50Y,2019ApJ...878...76Y}. Hence, they can be one of the possible sources of high-energy neutrinos detected by IceCube. Motivated by these studies and the lack of understanding about the exact source of the IceCube detected high-energy neutrinos, we conduct the first study on whether galaxy mergers can be the sources of these high-energy astrophysical neutrinos. To quantify this, we analyze whether there is a significant correlation between galaxy merger data obtained from various observational surveys and neutrino event data from IceCube. Specifically, we work with six different catalogs of galaxy mergers compiled by using observations of various surveys~ \cite{2009ApJS..181..233H,2018MNRAS.479..415A,2004AJ....128...62G,2005AJ....130.2043P, 2023MNRAS.524.4482B,2008MNRAS.388.1537M}. We perform the search analysis utilizing the method of unbinned maximum-likelihood-ratio~\cite{Braun:2008bg, Braun:2009wp, Abbasi:2010rd, Aartsen:2013uuv} and the publicly available 10-year data of muon-track events released by IceCube~\cite{Abbasi:2021bvk,website_neutrino_data}. We have performed both single source and stacking analysis. In the single source analysis, we have found that none of the galaxy mergers from all these catalogs is significantly correlated with IceCube neutrino events. In stacking analysis, we have worked with three weighting schemes and found the upper limits (at $95\%$ CL) of the all flavor high-energy neutrino flux,associated with galaxy mergers under consideration, at $100$ TeV with spectral index $\Gamma=-2$ are $2.57\times 10^{-18}$, $8.51 \times 10^{-19}$ and $2.36  \times 10^{-18}$ $\rm GeV^{-1}\,cm^2\,s^{-1}\,sr^{-1}$. In conclusion, these galaxy mergers do not contribute significantly to the IceCube detected high energy neutrino flux. However, we are hopeful that the acquisition of more data on galaxy mergers holds promise for illuminating the possibility of these mergers as sources of high-energy neutrinos. Furthermore, the IceCube-Gen2 ~\cite{IceCube-Gen2:2020qha} and KM3NeT~\cite{KM3NeT:2018wnd} telescope can also provide an opportunity to understand the origin of these neutrinos.

This paper is organized as follows. In Sec.~\ref{sec:nu_data}, we briefly overview the muon track datasets of high-energy neutrino events released by the IceCube Collaboration. In Sec.~\ref{sec:gal_merg}, we discuss the galaxy merger catalogs used in this work. In Sec.~\ref{sec_form}, we describe the unbinned maximum likelihood ratio method, which is used to test the possibility of galaxy mergers being the sources of high-energy neutrinos detected at the IceCube. In Sec.~\ref{sec:results}, we present the results of our single source and stacking analysis. Finally, we conclude this work in Sec.~\ref{sec:conc}.

\section{IceCube Neutrino Data}\label{sec:nu_data}
IceCube, located near the Amundsen–Scott South Pole Station in Antarctica, is a cubic kilometre neutrino telescope that can observe the entire celestial sphere~\cite{IceCube:2006tjp}. IceCube comprises 5,160 digital optical modules (DOMs) installed on 86 vertical strings from depths of 1,450 to 2,450 meters. These strings are positioned in a hexagonal grid with a horizontal spacing of 125 meters. Each string has 60 DOM placed at a vertical separation of 17 meters~\cite{IceCube:2008qbc,IceCube:2013dkx,IceCube:2016zyt}.

IceCube has two main types of neutrino signatures:\,\,cascades and tracks. When a very high energy (electron, muon, or tau) neutrino or antineutrino traverses through the pure Antarctic ice or the surrounding Earth's material, it can interact with the constituent nuclei through deep inelastic interaction by the exchange of a W or a Z boson. W-boson exchange (charged-current interaction) produces associated charged lepton $\nu_\ell \,+\, N \rightarrow \ell^- \,+\, N'$, where $\nu_\ell$  represents the neutrino associated with $\ell$ lepton, $N$ is the nucleon of the constituent matter through which the neutrino is traversing, and $N'$ is the final set of hadrons. The Z-boson exchange (neutral-current interaction) keeps the initial neutrino flavor intact, $\nu_\ell \,+\, N \rightarrow \nu_\ell \,+\, N'$. The muon, produced by the $\nu_\mu$ interaction, can travel a long distance, whereas an electron produced from a $\nu_e$ interaction rapidly loses energy due to its low mass. The final-state hadrons do not travel for a long distance beyond their production point. The tau lepton can produce a more complicated morphology depending on its energy {~\cite{Learned:1994wg,IceCube:2015vkp,2017arXiv170205238X,2018PhRvL.120x1105K,2018arXiv181201036V,2019arXiv190905162W,Stachurska:2019wfb,IceCube:2020fpi,Tian:2021jyy,2022arXiv220313827A,IceCube:2024pov}. Since muons can travel kilometres from their production point, they can enter the instrumented volume even if produced outside the detector volume. Electrons, hadrons, and taus need to be typically produced inside IceCube to be detected.
	
In the TeV-PeV energy range, other interactions are also present in addition to the above-mentioned processes. For instance, a high-energy electron antineutrino can interact with an electron at the antineutrino energy of $6.3$ PeV in the rest frame of the electron to produce an on-shell W boson in resonance, $\bar{\nu_e}\,+\,e^{-}\rightarrow W^{-}$,  called Glashow resonance~\cite{Glashow:1960zz,IceCube:2021rpz}. Additionally, a neutrino can interact with the target nucleus via a virtual photon to produce an on-shell W boson ($\nu_\ell\,+\,\gamma^*\rightarrow \ell^{-}\,+\, W^{+}$) and also trident production ($\nu\,+\,\gamma^{*}\rightarrow \nu\,+\,\ell_1^{-}\,+\ell_2^{+}$)~\cite{Seckel:1997kk,Alikhanov:2015kla,Zhou:2019vxt,Zhou:2019frk} can take place. Moreover, a recent study by Plestid and Zhou~\cite{Plestid:2024bva} showed that the final state radiation process, $\nu_\ell\,+\, N\rightarrow \,\ell\,+X+\gamma$, from neutrino interactions can have an important impact in the neutrino energy reconstruction. 

The highly energetic charged leptons travel faster than light in ice and produce Cherenkov light. The DOMs deployed in the strings detect the bluish Cherenkov light produced inside the detector volume. High-energy electrons lose energy rapidly in matter, making a spherical blob signature (through electromagnetic shower) called cascades. Hadrons can also produce cascades through hadronic showers. Tau neutrino creates a spatially separated double cascade signature. Due to the long lifetime, muon travels a long distance to create an elongated tracklike signature in IceCube. Track events have an excellent angular resolution ($\lesssim 1^\circ$) compared to cascade events ($\lesssim 10^\circ$), and hence they are well suited for astrophysical point source searches. 

IceCube published its 10-year track events data observed from April 6, 2008, to July 8, 2018, which consists of through-going events from all directions as well as events starting within the detector volume~\cite{Abbasi:2021bvk,website_neutrino_data}. There are, in total, 1,134,450 events in the data set. Later, Ref.\,\cite{2022PhRvD.105i3005Z} found 19 dimuon events in the dataset, but IceCube declared those were doubly counted events. In this work, we use the corrected dataset by Zhou and Chang, in which they exclude these 19 double-counted events. This dataset is publicly available in a Github repository{\footnote{\url{https://github.com/beizhouphys/IceCube_data_2008--2018_double_counting_corrected}}. The 10-year data is classified into five samples associated with different construction phases of IceCube:\,\,{\it (i)} IC40, {\it (ii)} IC59, {\it (iii)} IC79, {\it (iv)} IC86-I and {\it(v)} IC86-II to IC86-VII~\cite{2011ApJ...732...18A,IceCube:2013kvf,IceCube:2014vjc,IceCube:2019cia}. Here, the numbers stand for the number of strings in the IceCube construction phase.

\section{Galaxy merger Catalog}\label{sec:gal_merg}
Galaxy mergers are an inevitable prediction of $\Lambda$CDM cosmology and are the formation channel for massive galaxies. Cosmological simulations~\cite{2020MNRAS.494.4969P,2024arXiv240217889P} are used to study galaxy interactions and mergers over extended periods. Ferreira et al.\,\cite{2024MNRAS.533.2547F} have published a catalog of merging galaxies based on Ultraviolet Near-Infrared Optical Northern Survey (UNIONS) r-band images. Their work utilized a simulation-driven CNN model trained on mock observations from IllustrisTNG. Recently, the James Webb Space Telescope (JWST) has unveiled a hidden population of high-redshift galaxies, prompting numerous studies on galaxy merging events at these distances\,\cite{2024A&A...691A..69R,2024MNRAS.533.4472D,2024arXiv241104944D, Hsiao:2023nix, 2024ApJ...976L...8R,2024arXiv240709472D}. MACS0647-JD is likely the most distant known galaxy merger, observed by JWST located at a redshift of $z$ = 10.17\,\cite{Hsiao:2023nix}. Furthermore, dual active galactic nuclei and strongly lensed systems have been identified using MUSE adaptive-optics-assisted spectroscopy\,\cite{2024A&A...690A..57S}. It has been suggested that galaxy mergers can accelerate cosmic ray particles to ultrahigh energies.  References \cite{1993ICRC....2..341C,Jones:1997qr,2010A&A...524A..27L} investigated that the first-order Fermi acceleration mechanism within magnetized halos containing ionized gas associated with the merging galaxies can produce ultrahigh energy cosmic rays (UHECRs) up to $\sim 10^{20} \rm eV$. Later, Kashiyama and M\'{e}sz\'{a}ros~\cite{2014ApJ...790L..14K} suggested that a diffusive shock acceleration mechanism can also produce UHECRs with energies up to 0.1-1 EeV in merging galaxies. However, UHECRs lose their energy through hadronuclear interactions within the dynamic timescale of the shock. The $pp$ interactions during this process produce charged pions with the same multiplicity, which subsequently decay into very high-energy neutrinos. In Fig.~\ref{fig:cartoon_merger}, we present a schematic picture of the high-energy neutrino production process in a galaxy merger. According to their model, these neutrinos can contribute a significant portion of the neutrino flux observed by IceCube. Similarly, Refs.~\cite{2018ApJ...857...50Y, 2019ApJ...878...76Y} have also explored the production of high-energy neutrinos within galaxy mergers and suggested that they can be potential candidates for high-energy astrophysical neutrinos. These theoretical works motivate us to investigate the possibility of galaxy mergers being the source of high-energy neutrino events detected by IceCube. To perform this search analysis, we collect the data of galaxy mergers observed by various astronomical surveys. Specifically, we have taken the galaxy merger data from six catalogs in this work. \\
\begin{figure}[t!]
	\centering
	\includegraphics[width=.48\textwidth]{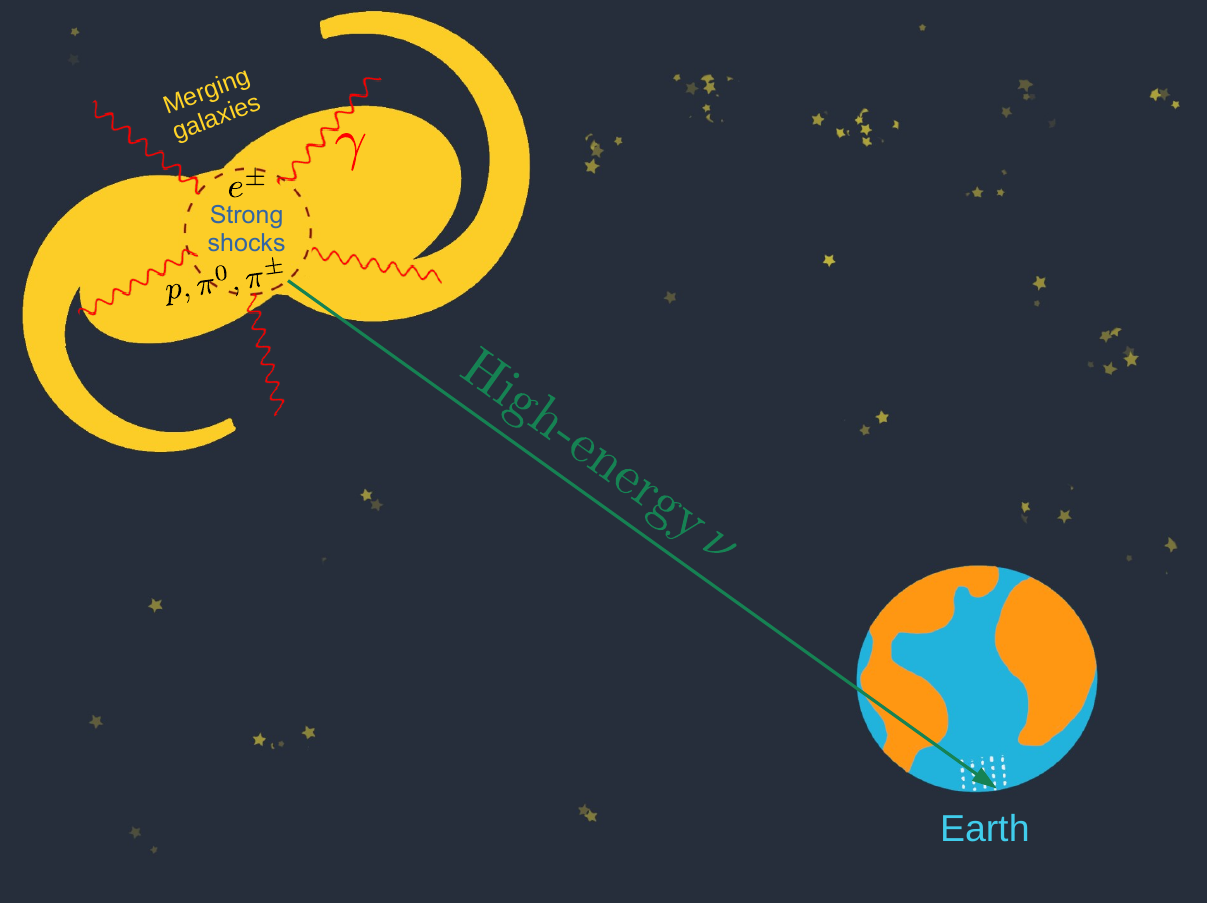}
	\caption{Illustration depicting the mechanism behind the generation of high-energy neutrinos originating from a galaxy merger event. Figure not to scale.}\label{fig:cartoon_merger}
\end{figure}
\begin{figure*}[t!]
	\centering
	\includegraphics[scale=0.68]{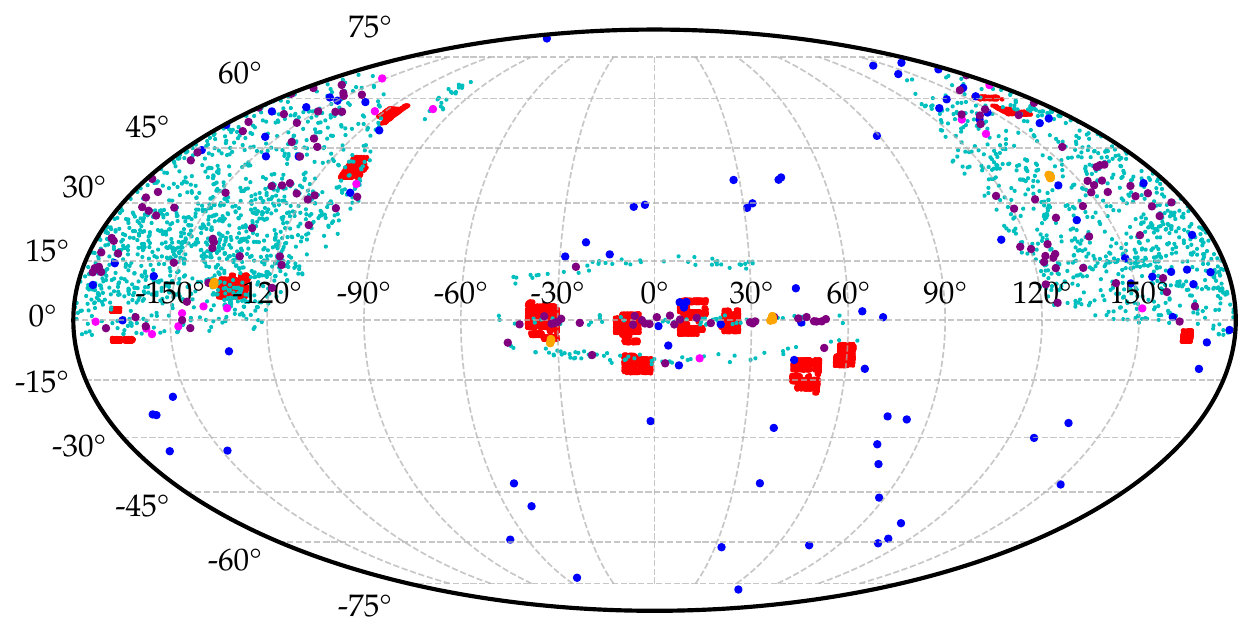}
	\caption{Locations of galaxy mergers from six different catalogs considered in this analysis are shown using different colors on a sphere with an equatorial coordinate system, which uses declination (vertical lines) and right ascension (horizontal lines)  to specify a point on the sphere. The points in red, cyan, orange, blue, purple and magenta colors represent the Galaxy mergers from catalogs: J/ApJS/181/233~\cite{2009ApJS..181..233H}, J/MNRAS/479/415~\cite{2018MNRAS.479..415A}, J/AJ/130/2043~\cite{2005AJ....130.2043P}, J/AJ/128/62~\cite{2004AJ....128...62G}, Dual AGN catalog~\cite{2023MNRAS.524.4482B}, and J/MNRAS/388/1537~\cite{2008MNRAS.388.1537M}, respectively (please see text for more details on these catalogs).}\label{fig:source_loc}
\end{figure*}
{\it (i)}  We choose 13,971 merger samples from the catalog J/ApJS/181/233~\cite{2009ApJS..181..233H}. Here, Hwang and Chang utilized observations from Red Sequence Cluster Survey 2 spanning the broad field of 422 square degrees in the sky using the Canada–France–Hawaii Telescope. They implemented pattern recognition algorithms to find merging galaxies. We have selected sources labeled ``type M'' from this catalog, characterized by closely merged objects with tails and deformed shapes. Objects mentioned in the catalog are up to redshift $z \sim$ 1. In Fig.~\ref{fig:source_loc}, we show the locations of these mergers by red color points on a sphere with an equatorial coordinate system. \\
{\it (ii)} We use 2,244 galaxy mergers from the catalog J/MNRAS/479/415~\cite{2018MNRAS.479..415A} as detailed in work by Ackermann et al., where deep convolution networks are implemented to detect mergers. They used a four-fold cross-validation method based on transfer learning methods to pick up the merger objects. The galaxy mergers are in $0.01 \lesssim z \lesssim 1.3469$. In Fig.~\ref{fig:source_loc}, we show the locations of these mergers by cyan color points on a sphere with an equatorial coordinate system.  \\
{\it (iii)}  The catalog J/AJ/130/2043 taken from the Canadian Network for Observational Cosmology field galaxy redshift survey consists of 70 close pairs of galaxies. This survey encompasses four distinct sections of the sky, spanning 1.5 square degrees. The details of the catalog can be found in the study by Patton \textit{et al.}~\cite{2005AJ....130.2043P}. The galaxies within this catalog have redshifts $0.1 \lesssim z \lesssim 0.6$. In Fig.~\ref{fig:source_loc}, we show the locations of these mergers by orange color points on a sphere with an equatorial coordinate system. \\
{\it (iv)}  We consider 107 double nucleus disk galaxies (DNDG) from the catalog  J/AJ/128/62 as documented in Gimeno \textit{et al.}~\cite{2004AJ....128...62G}. DNDG are indicators of a previous or ongoing merger and are crucial in understanding galaxy interactions, starburst and nuclear activity generation processes. The candidates are in the redshift range of $0.00328 \lesssim z \lesssim 0.05435$. We calculated the angular extensions, assuming a radius of 50 kpc for the nearest few DNDGs. We found that 11 candidates have angular extensions greater than 0.2$^\circ$, the lowest angular uncertainty in the reconstructed directions of neutrino events observed at IceCube. Hence, we have excluded these DNDGs from our analysis. In Fig.~\ref{fig:source_loc}, we show the locations of these mergers by blue color points on a sphere with an equatorial coordinate system.  \\
{\it (v)}  We incorporate 159 dual active galactic nuclei lying in the redshift range of $z$ = 0.023 to 0.26. The specifications of the objects can be found in Ref.~\cite{2023MNRAS.524.4482B}. The authors have used an automated search called graph-boosted iterated hill climbing to identify the candidates within 1 million SDSS DR16 galaxies. We call this group of candidates a dual AGN catalog. In Fig.~\ref{fig:source_loc}, we show the locations of these mergers by purple color points on a sphere with an equatorial coordinate system.  \\
{\it (vi)} Finally, we pick 21 progenitors of massive central merger systems from the catalog J/MNRAS/388/1537 with $0.027 \lesssim z \lesssim 0.12 $. The details can be found in Ref.~\cite{2008MNRAS.388.1537M}. The progenitors are believed to produce $\mathrm{M}_\mathrm{star} \gtrsim 10^{11}\mathrm{M}_\odot$ galaxies in large groups and clusters in future. In Fig.~\ref{fig:source_loc}, we show the locations of these mergers by magenta color points on a sphere with an equatorial coordinate system. 

We identified 23 galaxy mergers common across the catalogs and ensured that each source was considered only once during the analysis. Additionally, we examine a few interesting galaxy mergers  that are of particular interest since they have been theoretically modeled to study their high-energy astrophysical emissions: UGC 12914/5, UGC 813/6, VV 114 (discussed in Ref.~\cite{1993AJ....105.1730C,2002AJ....123.1881C,2014ApJ...790L..14K,2001A&AT...20..717V}), NGC 660, and NGC 3256 (mentioned in Ref.~\cite{2019ApJ...878...76Y}). Furthermore, we explored ESO 303-IG 011, a pair of star-forming galaxies captured by the Hubble Space Telescope and SDSS J084905.51+111447.2, a system of three colliding galaxies revealed by Chandra data. These candidates have some interesting common characteristics: \textit{(i)} all the candidates are close mergers (UGC 12914/5, UGC 813/6, ESO 303-11, VV 114) or appear to have merged already so that their nuclei lie within a massive bulge in the center of the galaxy (NGC 660, NGC 3256). \textit{(ii)} The few galaxies that are separated (UGC 12914/5, UGC 813/6, ES0 303-11, VV 114) appear to be major mergers, i.e., their galaxy or bulge masses are similar. \textit{(iii)} All the seven galaxies have extended tidal tails, bridges or loops of hot, magnetized gas that can be clearly distinguished in the blue (or g) band images. For example, the ``Taffy” galaxies (UGC 12914/5, UGC 813/6)~\cite{1993AJ....105.1730C,2002AJ....123.1881C} have bridges of neutral and ionized gas connecting the nuclei. The gas gives out synchrotron emission, which is detected in radio observations. Also, in the case of ESO 303-11, there is a loop of hot gas connecting the two nuclei, and it also appears to be forming stars. 

\section{Formalism}\label{sec_form}
 In this work, we employ the unbinned maximum likelihood ratio method to search for sources of high-energy astrophysical neutrinos detected by the IceCube Neutrino Observatory. This method, well described in the literature~\cite{Braun:2008bg, Braun:2009wp, Abbasi:2010rd, Aartsen:2013uuv}, has been used to search for various proposed sources~\cite{Abbasi:2010rd, Aartsen:2013uuv, Aartsen:2016lir, Hooper:2018wyk, Aartsen:2020eof, Smith:2020oac} and has proven to be an effective tool in the quest for neutrino source search analysis. 
 
We provide a brief overview of the general procedure for utilizing this method. We follow ref.\,~\cite{Zhou:2021rhl} closely in this analysis. The 10-year data released by the IceCube collaboration comprises five distinct datasets, as discussed in the previous section. If the $k^{\rm th}$ dataset has $N_k$ events, and there are $n^k_s$ events associated with sources in this specific dataset, we can define the probability density of $i^{\rm th}$ event as follows:
\begin{equation}
	\frac{n_s^k}{N_k} S_{i}^k + \left(1-\frac{n_s^k}{N_k}\right) B_i^k  \,,
	\label{eq:ev-prob}
\end{equation}
where the first and second term corresponds to the relative contribution of the signal ($S^k_i$) and background ($B^k_i$) probability density functions (PDFs) of $i^{\rm th}$ event in $k^{\rm th}$ dataset. The likelihood function, denoted as $\mathcal{L} \left(n_s \right)$, is defined as the product of the probability densities of all neutrino events within the 10-year dataset and can be expressed as:
\begin{equation}
	\mathcal{L} \left(n_s \right) = \prod_{k} \prod_{i \in k}  \left[ \frac{n_s^k}{N_k} S_{i}^k + \left(1-\frac{n_s^k}{N_k}\right) B_i^k \right] \,,
	\label{eq:lik_func}
\end{equation}
where index $k$ ranges from $1$ to $5$, covering all five datasets, while $i$ iterates over all the events within the $k^{\rm th}$ dataset. The parameter $n_s$ represents the total number of events associated with sources from all five datasets, an unknown parameter we aim to determine by maximizing the likelihood function. Furthermore, $n_s$ is related to $n^k_s$ as $n^k_s = f_k \times n_s$ with $f_k$ being the expected fractional contribution to the events from the $k^{\rm th}$ dataset.  

We will perform two different types of analysis:\\
1. single source analysis and \\
2. stacking analysis. \\
In the case of single source analysis, $f_k$ can be computed as 
 \begin{equation}
	f_k = 
	\frac{ w^k_{j, \rm acc} }{\sum_{k} w^k_{j, \rm acc}} ,
	\label{eq_fk_1}
\end{equation}
and for stacking analysis, as
\begin{equation}
	f_k = 
	\frac{ \sum_{j} w_{j, \rm model} w^k_{j, \rm acc} }{\sum_{k} \sum_{j} w_{j,\rm model} w^k_{j, \rm acc}} \,.
	\label{eq_fk_2}
\end{equation}
Here, $w^k_{j, \rm acc}$ is the weighting factor, which depends on the detector's response and is defined as
\begin{equation}
	w^k_{j, \rm acc} (\delta_j)
	\propto t_k \times \int A^k_\text{eff}(E_\nu, \delta_j) E_\nu^{\Gamma} \, d E_\nu \,,
\end{equation}
with $A^k_\text{eff} (E_\nu, \delta_j)$ and $t_k$ being the effective area of the IceCube detector and total uptime for the $k^{\rm th}$ dataset, respectively.  The source declination and neutrino energy are represented by $\delta_j$ and $E_\nu$,  respectively. We have assumed that the neutrino energy spectrum at the source is a power law with $\Gamma$ being the spectral index.  The weighting factor, that depends on the intrinsic property of sources, is denoted by $w_{j, \rm model}$. We explore three different weighting schemes to fix $w_{j, \rm model}$:
\begin{enumerate}
	\item For the first case, the $w_{j, \rm model}$ is assumed to be independent of the source's intrinsic properties. We call this uniform weighting. Therefore, we have $w_{j, \rm model} =1$. 
	\item Since the galaxy mergers used in the work are distributed over a range of redshifts, in the second case, we choose the weighting factor
	\begin{equation}
		w_{j, \rm model} \propto 1/{d_L(z)}^2\,,
	\end{equation}
	where $d_L(z)$ is the luminosity distance to a source located at redshift $z$. We call this the luminosity distance weighting scheme.  
	
	\item As discussed in Sec.~\ref{sec:gal_merg}, the neutrinos are expected to originate in galaxy mergers via the decay of charged pions, which are produced through interactions between cosmic rays, predominantly protons, and target gas nuclei ($pp$) or photons ($p\gamma$). Yuan \textit{et al.}~\cite{2019ApJ...878...76Y} studied the redshift dependence of high-energy neutrino production in galaxy mergers. Their study showed that the neutrino flux depends on the cosmic ray energy input rate, galactic radius, shock velocity, and galactic magnetic field, which are all redshift-dependent. To include such redshift dependence of high-energy neutrino flux, we choose the weighting factor as~\cite{2019ApJ...878...76Y}:
		\begin{equation}
			{
				w_{j, \rm model} \propto  \frac{\epsilon_\nu Q_\nu^{(g)}}{(1+z)^2 H(z)}\,,}
		\end{equation}
		where $ \epsilon_\nu Q_\nu^{(g)}$ is the all-flavor neutrino luminosity density in galaxy merger and $ H(z) $ is the hubble parameter. Neutrino luminosity density depends on the $pp$ optical depth and cosmic ray energy input rate. In our study, we took the cosmic ray energy input rate as a function of $z$ from Ref.~\cite{2019ApJ...878...76Y}. Specifically, we took the magenta dashed curve from Fig. 3 of Ref.~\cite{2019ApJ...878...76Y} as a representative redshift dependence of cosmic ray energy input rate. We call this the Yuan \textit{et al.} weighting scheme.

\end{enumerate}

We will now define the signal and background PDFs. The signal PDF for an $i^{\rm th}$ event and a source located at coordinates $\vec{x}_i$ and $\vec{x}_j$, respectively, is represented by a Gaussian PDF as
\begin{equation}
	S_{ij}^k \left( \vec{x}_i, \sigma_i, \vec{x}_j \right) 
	= 
	\frac{1}{2 \pi \sigma_i^{2}} \exp\left( -\frac{ D( \vec{x}_i, \vec{x}_j )^2 } {2 \sigma_i^{2} }\right)\, ,
	\label{eq:Gaussian}
\end{equation}
where $\sigma_i$ and $D( \vec{x}_i, \vec{x}_j )$ are the uncertainty in event direction and the angular distance between an event and source, respectively. In case of single source analysis, $S^k_i \equiv S_{ij}^k \left( \vec{x}_i, \sigma_i, \vec{x}_j \right)$, as $j$ is fixed for each source. In analysis with multiple sources, $S^k_i$ is given as
\begin{equation}
	S_i^k
	= 
	\frac{\sum_{j} w_{j, \rm model} w^k_{j, \rm acc} S_{ij}^k}{\sum_{j} w_{j,\rm model} w^k_{j, \rm acc}}\, .
	\label{eq:spdf_stack}
\end{equation} 
Let us now define the background PDFs. The major contribution in the background comes from the atmospheric neutrino data. The variation of the atmospheric neutrino is very small over a small region. Therefore, in a small region, we can assume the background PDF is uniform. For single source analysis, we choose the background PDF as
\begin{equation}
	B_i(\delta_i)=\frac{1}{\Omega_{\delta_{src}\pm 3 ^\circ}}\, ,
\end{equation}
 where $\Omega_{\delta_{src}\pm 3 ^\circ}$ is the right ascension integrated solid angle in the $\delta_{src} \pm 3 ^\circ$ region with $\delta_{src}$ and $\delta_i$ denoting the declination of point source and neutrino event. Similarly, we have chosen a region of interest (ROI) consisting of a declination band spanning $\pm 3^\circ$ centred on the declination of the source to compute the signal PDF in a single-source analysis. Only neutrino events within the ROI are considered for the single source analysis.

 In the case of stacking analysis, we derive the background PDF using the IceCube neutrino data directly using the following expression~\cite{Li:2022vsb}:
  \begin{equation} 
  	B_i(\delta_i)=\frac{N_{\delta_i\pm3^\circ}}{N_{tot} \Omega_{\delta_i\pm 3 ^\circ}}\, ,
  \end{equation} 
where $N_{tot} $ is the total number of neutrino events in a specific dataset and $N_{\delta_i\pm3^\circ}$ is the number of neutrino events in the $\delta_i\pm 3 ^\circ $ region; $\Omega_{\delta_i\pm 3 ^\circ}$ is the solid angle corresponding to the $\delta_i \pm 3^ \circ $ region.

We define a test statistic (TS) to quantify the compatibility of IceCube data with background or signal events as,
\begin{equation}
	\text{TS}(n_s) = - 2 \ln{ \frac{ \mathcal{L}(n_s=0) }{ \mathcal{L}(n_s) } }\,.
	\label{eq_TS}
\end{equation}
Here, $n_s = 0$ is the null hypothesis, which states that all observed events solely consist of background. Large test statistic values will imply a reduced compatibility with the null hypothesis. We, therefore, maximize the test statistic using MINUIT~\cite{James:1975dr,James:1994vla,iminuit} to obtain the best fit $n_s$ values and corresponding test statistic (TS$_{\rm max}$). Wilk's thorem~\cite{10.1214/aoms/1177732360} in the context of such likelihood test ratio analysis gives a very important result that the TS$_{\rm max}$ will asymptotically follow a $\chi^2$ distribution with degrees of freedom equals to the difference between the number of free parameters in two hypotheses if the null hypothesis is favored. We utilize this result in our single source analysis to quantify the compatibility between observed events and signals from galaxy mergers. In stacking analysis, wherein the contributions of all the sources are summed up to find the significance, we vary the Test Statistics as a function of the flux normalization of muon neutrino at $100$ TeV to assess the significance of any possible signal over the background. We will discuss this in detail in the next section. 
\begin{figure}[t]
	\centering
	\includegraphics[width=.48\textwidth]{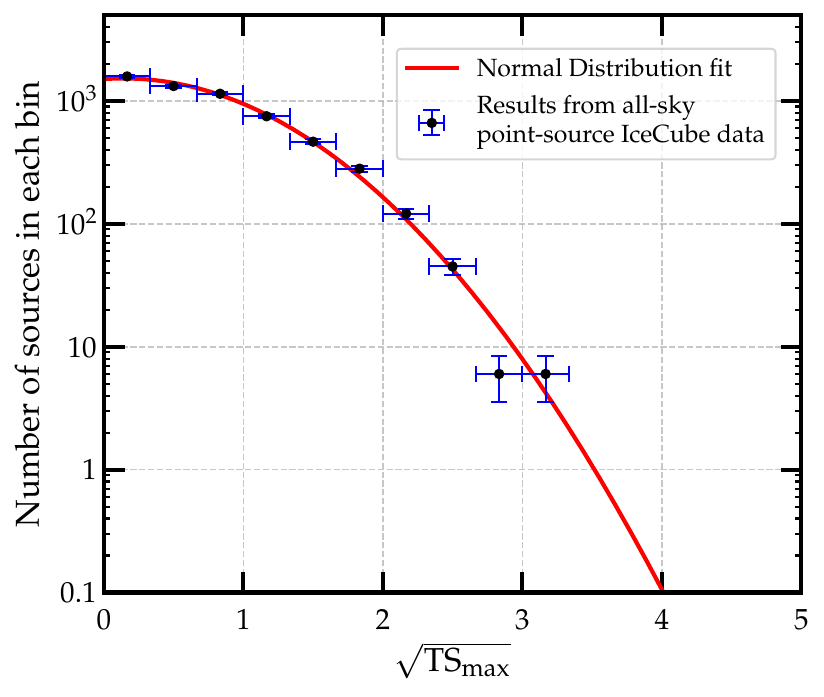}
	\caption{Blue data points with error bars represent the number of sources in a given bin of $\sqrt{\rm TS_{\rm max}}$ obtained in our single source analysis. The error bars are obtained by taking the square root of the number of sources in each bin. The red curve represents the best-fit distribution.}\label{fig:TSm_g2}
\end{figure}

\begin{figure*}[htbp]
	\centering
	\includegraphics[scale=0.6]{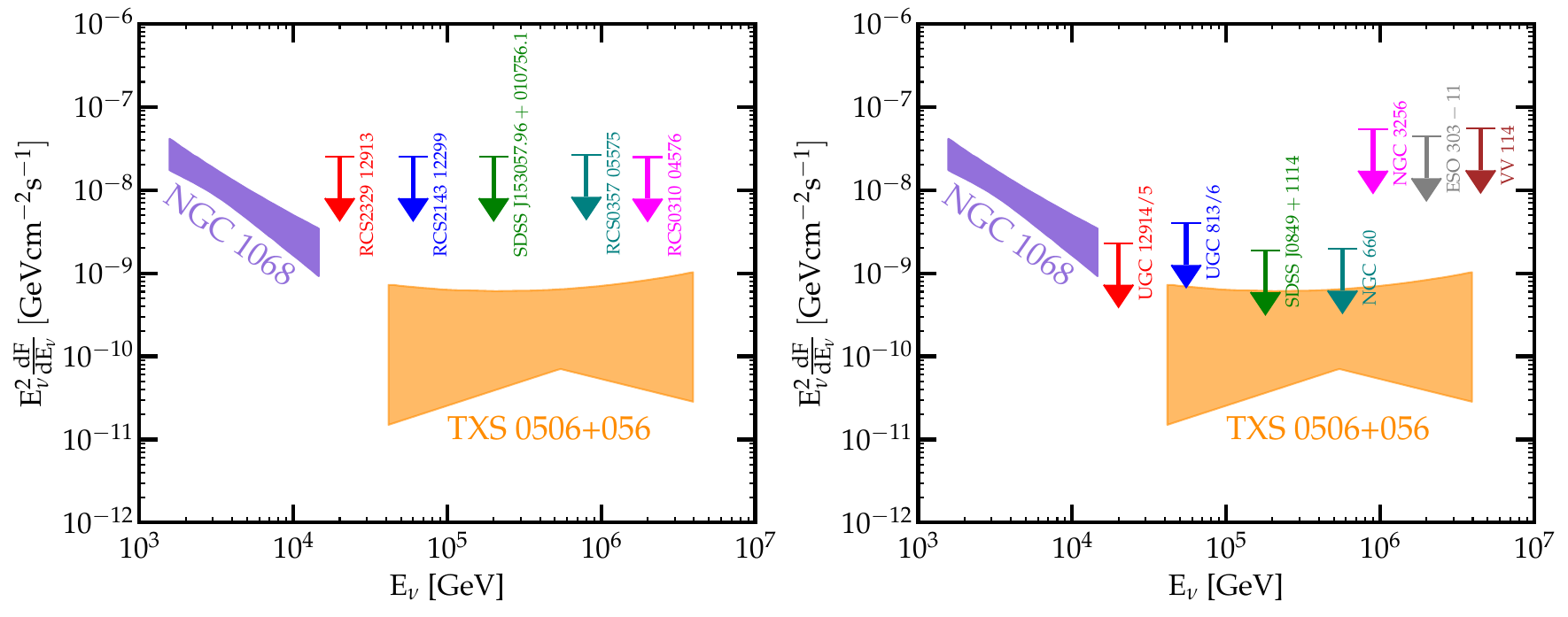}
	\caption{Upper limits on the total flux of high-energy neutrinos with $\Gamma=-2$ at various energies for the five most significant sources among all the considered galaxy mergers within six catalogs ({\it left-hand panel}) and seven interesting galaxy mergers mentioned in Sec.~\ref{sec:gal_merg} ({\it right-hand panel}) are shown. The purple and orange shaded regions show the observed flux from NGC 1068~\cite{Aartsen:2019fau,IceCube:2022der} and TXS 0506+056~\cite{IceCube:2018cha, IceCube:2018dnn}, respectively.}\label{fig:sing-flux}
\end{figure*}

\section{Results}\label{sec:results}
In this section, we present the results of our analysis involving galaxy mergers as a possible source of high-energy astrophysical neutrino events observed by IceCube. 

\begin{table*}[htbp]
	
	\fontsize{8pt}{10pt}\selectfont
	
	\begin{tabular}{|c|c|c|c|c|c|}
		\hline
		Identity & (RA, DEC) & $n_{s}$, TS$_{\rm max}$ & Pre-trial p value   & Post-trial p value & Upper limit on total flux\\
			 &  &  & and significance & and significance & ($\rm GeV^{-1}cm^{-2}s^{-1}$)\\ \hline  
			\begin{tabular}{@{}c@{}c@{}c@{}c@{}c@{}} RCS2329 12913\\  RCS2329 12918\\RCS2329 12920 \\RCS2329 12922\\RCS2329 12921\\RCS2329 12926\end{tabular}
		
		& \begin{tabular}{@{}c@{}c@{}c@{}c@{}c@{}}(347.74,-1.83)\\  (347.77,-1.65)
			\\(347.78,-1.54)\\(347.81,-1.93)\\(347.79,-1.46)\\(347.86,-1.51)\\
		 \end{tabular} & \begin{tabular}{@{}c@{}c@{}c@{}c@{}c@{}}    34.1, 10.25   \\ 33.6, 9.41\\32.2, 8.25	  \\ 30.5, 8.05\\30.4, 7.00\\29.9, 6.89\\
	   \end{tabular} &\begin{tabular}{@{}c@{}c@{}c@{}c@{}c@{}}$6.84\times 10^{-4}, 3.2\sigma$\\   $1.08\times 10^{-3}, 3.1 \sigma$   \\    $2.04\times 10^{-3}, 2.9\sigma$ \\ $2.27\times 10^{-3}, 2.8\sigma$\\ $4.08\times 10^{-3}, 2.6\sigma$\\ $4.33\times 10^{-3}, 2.6\sigma$ \end{tabular}  &\begin{tabular}{@{}c@{}c@{}c@{}c@{}c@{}}$1, 0\sigma$\\  $1, 0\sigma$  \\$1, 0\sigma$\\$1, 0\sigma$\\$1, 0\sigma$  \\$1, 0\sigma$  \\ \end{tabular} &\begin{tabular}{@{}c@{}c@{}c@{}c@{}c@{}}    $ \Phi_{3\nu} \leq 2.52\times 10^{-18}$   \\$ \Phi_{3\nu} \leq 2.48\times 10^{-18}$ \\$ \Phi_{3\nu} \leq 2.38\times 10^{-18}$\\ $ \Phi_{3\nu} \leq 2.25\times 10^{-18}$\\$ \Phi_{3\nu} \leq 2.25\times 10^{-18}$\\ $ \Phi_{3\nu} \leq 2.21\times 10^{-18}$ \end{tabular} \\ \hline

		\begin{tabular}{@{}c@{}c@{}c@{}c@{}c@{}} RCS2143 12299\\  RCS2143 12289\\RCS2143 12318 \\RCS2143 12284\\RCS2143 12370\\RCS2143 12342\end{tabular}
	
	& \begin{tabular}{@{}c@{}c@{}c@{}c@{}c@{}}(327.71,0.60)\\  (327.67,0.63)
		\\(327.79,0.58)\\(327.66,0.81)\\(328.01,-0.44)\\(327.88,-0.41)\\
	\end{tabular} & \begin{tabular}{@{}c@{}c@{}c@{}c@{}c@{}}    35.9, 9.77
	   \\ 35.9, 9.75\\35.0, 9.31	  \\ 34.2, 8.42
	   \\31.9, 8.05\\31.7, 7.80\\
	\end{tabular} &\begin{tabular}{@{}c@{}c@{}c@{}c@{}c@{}}$8.86\times 10^{-4}, 3.1\sigma$\\   $8.99\times 10^{-4}, 3.1 \sigma$   \\    $1.14\times 10^{-3}, 3.1\sigma$ \\ $1.86\times 10^{-3}, 2.9\sigma$ \\ $2.27\times 10^{-3}, 2.8\sigma$\\ $2.61\times 10^{-3}, 2.8\sigma$ \end{tabular}  &\begin{tabular}{@{}c@{}c@{}c@{}c@{}c@{}}$1, 0\sigma$\\  $1, 0\sigma$  \\$1, 0\sigma$\\$1, 0\sigma$\\$1, 0\sigma$  \\$1, 0\sigma$  \\ \end{tabular} &\begin{tabular}{@{}c@{}c@{}c@{}c@{}c@{}}    $ \Phi_{3\nu} \leq 2.52\times 10^{-18}$   \\$ \Phi_{3\nu} \leq 2.52\times 10^{-18}$ \\$ \Phi_{3\nu} \leq 2.45\times 10^{-18}$\\ $ \Phi_{3\nu} \leq 2.40\times 10^{-18}$\\$ \Phi_{3\nu} \leq 2.36\times 10^{-18}$\\ $ \Phi_{3\nu} \leq 2.35\times 10^{-18}$ \end{tabular} \\ \hline

	SDSS J153057.96+010756.1 & (232.74,1.13) & 37.4, 9.56
	 & $9.95\times 10^{-4}, 3.1\sigma$ & $1, 0\sigma$ & $ \Phi_{3\nu} \leq 2.62\times 10^{-18}$ \\ \hline

		\begin{tabular}{@{}c@{}c@{}c@{}c@{}c@{}c@{}}RCS0357 05575\\RCS0357 05549\\RCS0357 05603  \\ RCS0357 05596\\ \end{tabular}
		
		& \begin{tabular}{@{}c@{}c@{}@{}c@{}c@{}c@{}}(59.99,-7.84)\\  (59.89,-7.88)
			\\(60.19,-9.74)\\(60.14,-9.75)\end{tabular} & \begin{tabular}{@{}c@{}c@{}@{}c@{}c@{}}15.2, 8.11 \\ 15.0, 7.08 \\ 12.7, 6.83 \\ 12.6, 6.64   \end{tabular} &\begin{tabular}{@{}c@{}c@{}@{}c@{}c@{}c@{}}$2.20\times 10^{-3}, 2.8\sigma$\\   $3.90\times 10^{-3}, 2.7\sigma$   \\    $4.49\times 10^{-3}, 2.6\sigma$\\ $4.98\times 10^{-3}, 2.6\sigma$\\ \end{tabular}  &\begin{tabular}{@{}c@{}c@{}@{}c@{}c@{}c@{}}$1, 0\sigma$\\  $1, 0\sigma$  \\$1, 0\sigma$ \\$1, 0 \sigma$\\   \end{tabular} 	& \begin{tabular}{@{}c@{}c@{}@{}c@{}c@{}c@{}}$ \Phi_{3\nu} \leq 3.69\times 10^{-18}$ \\   $ \Phi_{3\nu} \leq 3.65\times 10^{-18}$ \\ $ \Phi_{3\nu} \leq 4.53\times 10^{-18}$ \\ $ \Phi_{3\nu} \leq 4.49\times 10^{-18}$  \end{tabular} \\ \hline	
		
			\begin{tabular}{@{}c@{}}RCS0310 04576 \\ RCS0310 04575\\\end{tabular}
		
		& \begin{tabular}{@{}c@{}}(49.81,-17.31)
			\\  (49.78,-17.37)\\ \end{tabular} & \begin{tabular}{@{}c@{}c@{}c@{}}  11.4, 7.03  \\  11.9, 6.89 \end{tabular} &\begin{tabular}{@{}c@{}c@{}}$4.00\times 10^{-3}, 2.7\sigma$\\   $4.33 \times 10^{-3}, 2.6 \sigma$   \end{tabular}  &\begin{tabular}{@{}c@{}c@{}}$1, 0\sigma$\\  $1, 0\sigma$  \\ \end{tabular}&  \begin{tabular}{@{}c@{}c@{}}$ \Phi_{3\nu} \leq 7.44\times 10^{-18}$\\  $ \Phi_{3\nu} \leq 7.78\times 10^{-18}$  \\ \end{tabular} \\ \hline

	\end{tabular}

	\centering
	\caption{Pre and post-trial p values, along with corresponding significance levels for some of the galaxy mergers with the highest significance in our single source analysis, are listed here. We have also presented the best fit $n_s$ and TS$_{\rm max}$ values for these galaxy mergers. Additionally, the upper limits on total neutrino flux associated with these mergers for $\Gamma=-2$ are listed.}
	\label{tab:p_values}
\end{table*}

\begin{table}[h]
	\begin{tabular}{|c|c| }
		\hline
		\multicolumn{2}{|c|}{Upper limit on total flux (in units of $\rm GeV^{-1}cm^{-2}s^{-1}$)} \\
		\hline
		~~~~~~~~~Source~~~~~~~~~ & Upper limit  \\
		\hline
		UGC 12914/5 &$ \Phi_{3\nu} \leq 2.300 \times 10^{-19} $\\
		UGC 813/6 &  $ \Phi_{3\nu}\leq 3.976 \times 10^{-19}$  \\
		SDSS J0849+1114 &$ \Phi_{3\nu} \leq 1.873 \times 10^{-19}$ \\
		NGC 660 &$ \Phi_{3\nu}\leq 1.955 \times 10^{-19} $\\
		NGC 3256 &  $ \Phi_{3\nu} \leq 5.408 \times 10^{-18}$  \\
		ESO 303-11 &$ \Phi_{3\nu}\leq 4.393 \times 10^{-18}$ \\ \hline
		VV 114 &$ \Phi_{3\nu} \leq 5.473 \times 10^{-18}$ \\
		\hline
	\end{tabular}
	\caption{ Upper limits on the flux at $100$ TeV ($\Phi_{3\nu}$) for seven interesting galaxy merger candidates listed in Sec.~\ref{sec:gal_merg} within our single source analysis are given. }
	\label{tab:upper_lim_th_source}
\end{table}

1. {\it Single source analysis:}
We calculated the test statistic for all galaxy mergers listed in the six catalogs, following the methodology outlined in the preceding section. Subsequently, we employed MINUIT to maximize the test statistic, obtaining the best-fit values for $n_s$ and the corresponding maximum TS value (TS$_{\rm max}$) for each source. To assess the distribution of
 sources with TS$_{\rm max}$, we plotted the number of sources against $\sqrt{\rm TS_{\rm max}}$ in fixed bins of $\sqrt{\rm TS_{\rm max}} = 0.33$, as depicted in the left-hand panel of Fig.~\ref{fig:TSm_g2}. Error bars in each bin were obtained using the square root of the respective number counts. We can see that the distribution exhibits a peak at near zero, indicating a preference for the background hypothesis. To quantitatively support this observation, we fitted the distribution, assuming a normal distribution, and found the best-fit values for mean, standard deviations and amplitude to be $0.139\pm 0.096$, $0.882 \pm 0.043$, and $1524.625\pm 73.716$, respectively. The best-fit values suggest that the distribution approximately resembles the standard normal distribution. Hence, according to Wilk's theorem, we cannot reject the null hypothesis. Consequently, we conclude that there is no significant correlation between the galaxy mergers listed in these catalogs and the high-energy neutrino events detected at IceCube.

Subsequently, we computed the local p-value (pre-trial)~\cite{Barlow:2019svl,Sinervo:2002sa,ParticleDataGroup:2022pth} and the corresponding significance level for each source. Additionally, we calculated the global p-value using the formula $p_{\rm global} = 1 - (1 - p_{\rm local})^{N}$, with $N$ representing the number of sources, to establish the post-trial significance. Table~\ref{tab:p_values} lists a few most significantly correlated sources from all these catalogs. In this table, we have grouped the mergers if the angular separation between any two of these mergers is within 1$^\circ$, as IceCube cannot distinguish them. Additionally, we have also computed the significance for seven interesting galaxy mergers, as mentioned in Sec.~\ref{sec:gal_merg}. Notably, we found that none of these sources exhibit a significant correlation with the high-energy neutrinos observed at IceCube. The source VV 114 was found to be the most significantly correlated among these sources with a pre-trial p-value and significance given by $2.12\times 10^{-2}, 2.03~\sigma$, respectively. However, it is important to note that while we did not find a significant correlation, the best-fit $n_s$ values allow us to set upper limits on the neutrino flux originating from these sources. To compute these upper limits, we used the following expression for the expected number of $\nu_\mu+ \bar{\nu}_\mu$ events at IceCube
\begin{equation}
	\begin{aligned}
		n_s =  2\pi \sum_{k} \, t_k  \int  d  \sin & \, \delta  \int A^k_\text{eff}(E_\nu, \delta) \,  \frac{d F}{d E_\nu}(E_\nu) \, d E_\nu \,,\label{eq:exp_nu}
	\end{aligned}
\end{equation}
where $ d F(E_\nu) / d E_\nu$ represents the flux of $\nu_\mu+ \bar{\nu}_\mu$. In this work we consider $d F(E_\nu) / d E_\nu=\Phi_0 \times \left({E_\nu}/{100\ \text{TeV}}\right)^{\Gamma}$ with $\Phi_0$ being the flux normalization of $\nu_\mu+ \bar{\nu}_\mu$ events at 100 TeV. We assume that neutrino flux is the same for all three neutrino flavors. Hence, we define the normalization of total neutrino flux at 100 TeV as $ \Phi_{3\nu}=3 \Phi_0 $. The computed upper limits are shown in Fig.~\ref{fig:sing-flux} for the five most significant sources and seven interesting sources mentioned in Sec.~\ref{sec:gal_merg}, assuming $\Gamma = -2$. In Fig.~\ref{fig:sing-flux}, we have also presented the best-fit flux for the two known sources of high-energy neutrinos, NGC 1068~\cite{Aartsen:2019fau,IceCube:2022der} and TXS 0506+056~\cite{IceCube:2018cha, IceCube:2018dnn} by the purple and orange shaded regions, respectively. It should be noted here that we found a nonzero $n_s \sim 8.3$ value only for VV 114 among the seven considered sources. Hence, we use the obtained $n_s$ to compute the upper limit for VV 114. To get the upper limits for the remaining six sources, we assume that $n_s$ follows the Poisson distributions. Under this assumption, we compute the upper limits (at 90$\%$ CL) on the neutrino flux from these six sources. The sources NGC 3256, ESO 303-11, and VV 114 are located in the southern hemisphere (with negative declination). Since the effective area for low-energy neutrinos is smaller in the southern hemisphere for IceCube, the neutrino flux needs to be higher to achieve the same number of expected neutrinos compared to sources located in the northern hemisphere. This explains why the upper limit of the total neutrino flux for these sources is higher than for the others. The computed upper limits for these seven sources are listed in Table~\ref{tab:upper_lim_th_source}.

\begin{figure*}[ht!]
	\centering
	\includegraphics[scale=0.6]{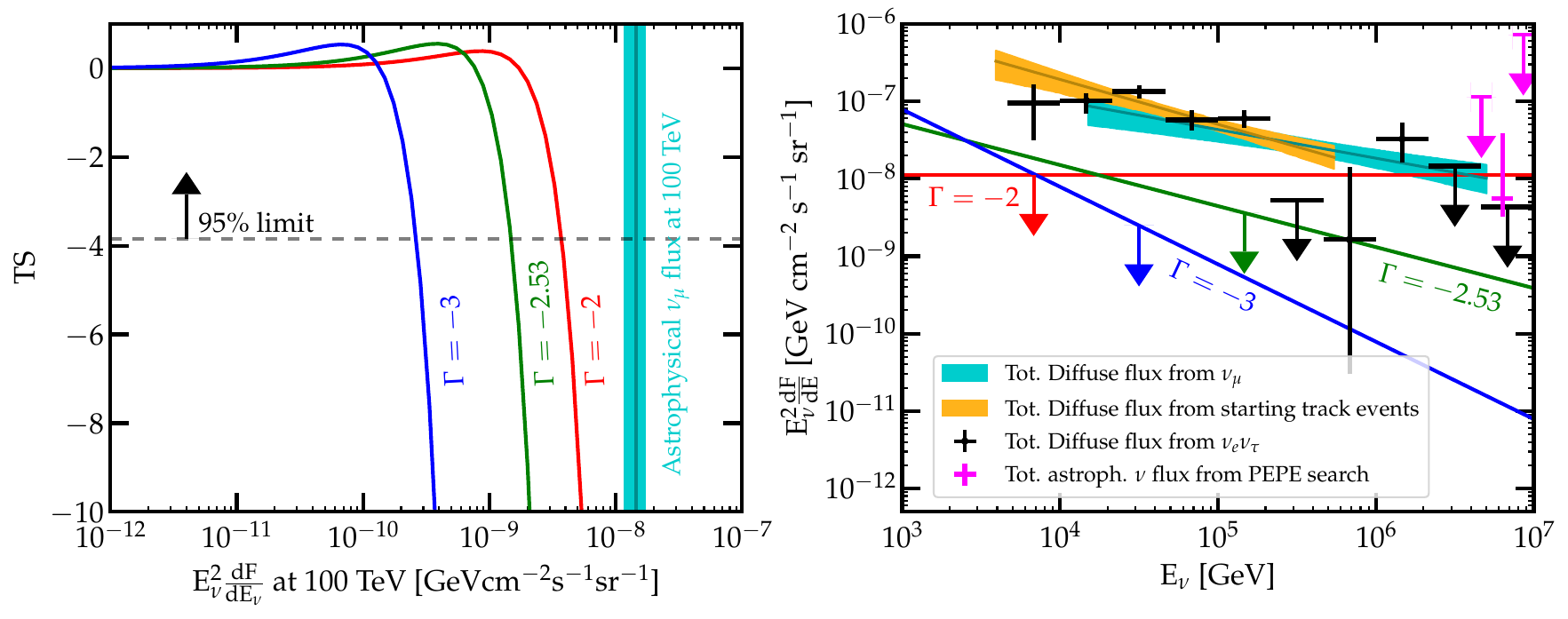}
	\caption{{\it Left-hand panel:} $\rm TS$ values obtained in our stacking analysis with all galaxy merger sources in the six catalogs within the uniform weighting scheme are plotted against $\Phi_0$ for $\Gamma= -2$ (red color), $-2.53$ (green color) and $-3$ (blue color). The black dashed line represents the $95 \%$ CL limit on $\rm TS$. The cyan band corresponds to the observed astrophysical muon neutrino flux by IceCube at $100$ TeV~\cite{IceCube:2021uhz}. {\it Right-hand panel:} The upper limits on the high-energy neutrino flux coming from all considered galaxy mergers in this work within the uniform weighting scheme for $\Gamma= -2$ (red color), $-2.53$ (green color) and $-3$ (blue color). The cyan-shaded and orange region shows the astrophysical diffuse high-energy neutrino flux measured using muon-neutrino and starting track events observed by IceCube, respectively~\cite{IceCube:2021uhz,Abbasi:2024jro}. The black data points with the error bar represent the total measured astrophysical diffuse high energy neutrino flux from combined electron and tau neutrino channels by IceCube~\cite{Aartsen:2020aqd}. We assume all the neutrino flavors are in equal ratio. The magenta points correspond to the measured astrophysical total neutrino flux from the PEPE analysis~\cite{IceCube:2021rpz}.
	}\label{fig:stack_1teV}
\end{figure*}

\begin{figure*}[ht!]
	\centering
	\includegraphics[scale=0.6]{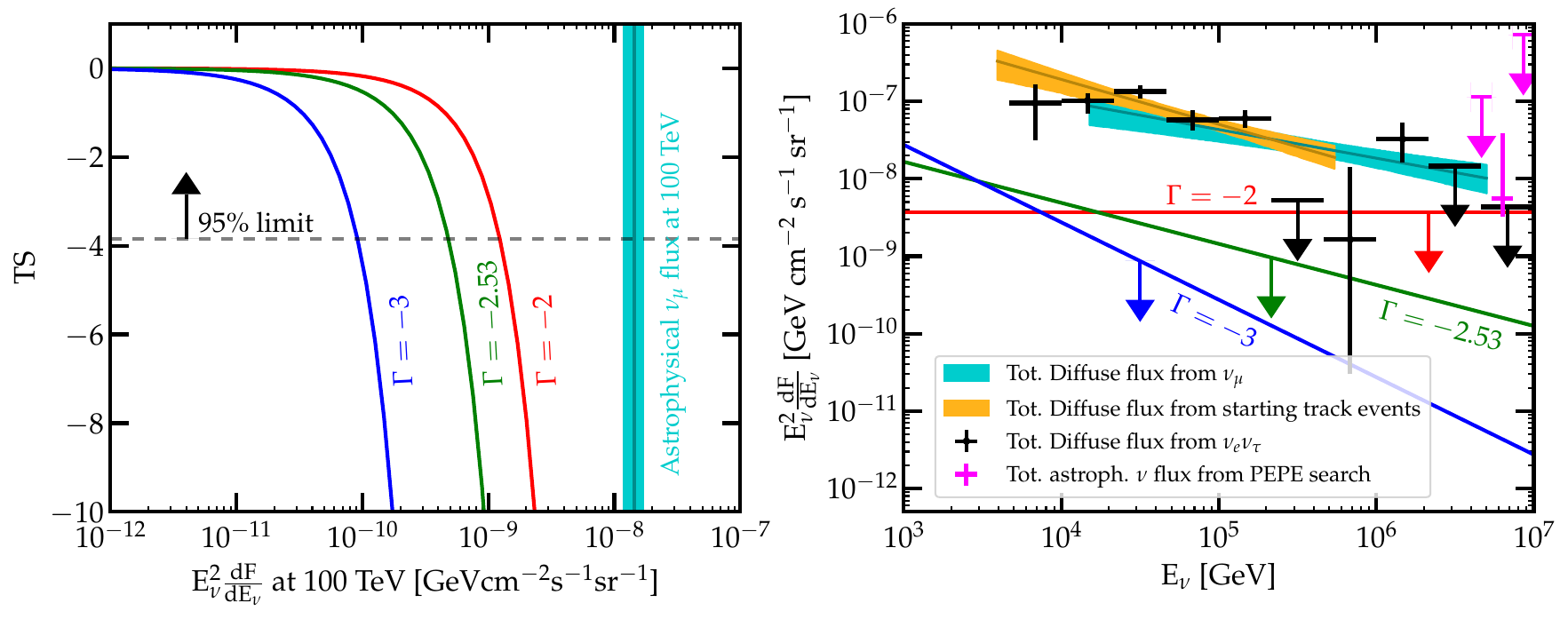}
	\caption{$\rm TS$ values ({\it left-hand panel}) and the upper limits on the all flavor high-energy neutrino flux ({\it right-hand panel}) in our stacking analysis with all considered galaxy merger sources within the luminosity distance weighting scheme are shown. We have chosen the same color and labeling schemes as considered in Fig.~\ref{fig:stack_1teV}.}\label{fig:stack_1teV_lum}
\end{figure*}

\begin{figure*}[ht!]
	\centering
	\includegraphics[scale=0.6]{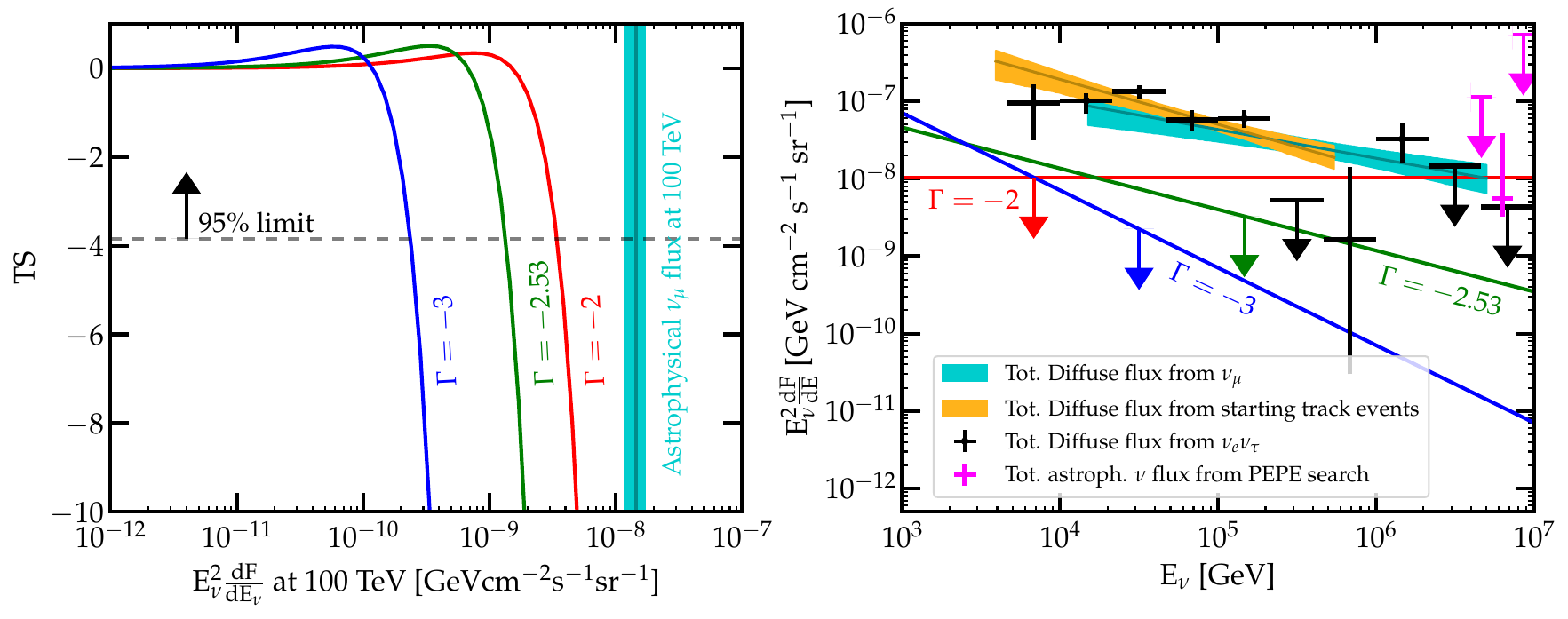}
	\caption{$\rm TS$ values ({\it left-hand panel}) and the upper limits on the all flavor high-energy neutrino flux ({\it right-hand panel}) in our stacking analysis with all considered galaxy merger sources within the Yuan \textit{et al.} weighting scheme are shown. We have chosen the same color and labeling schemes as considered in Fig.~\ref{fig:stack_1teV}.	}\label{fig:stack_1teV_w3}
\end{figure*}

2. {\it Stacking analysis:} In this section, we present the results of our stacking analysis. This method combines signals from different sources to estimate the expected number of detectable neutrino events at IceCube. We compute the expected events using Eq.~(\ref{eq:exp_nu}) for given fluxes by different sources. In this analysis, we assume that the energy dependence of the neutrino flux from these sources can be described by a power-law, and we consider the flux to be the same for all three types of neutrino flavors. We explore three cases with different spectral indices: $\Gamma = -2.0, -2.53,$ and $-3.0$. We choose these values for specific reasons. The first choice, $\Gamma = -2.0$, is based on the Fermi acceleration mechanism, which predicts a similar-tilted spectrum. The second choice, $\Gamma = -2.53$, is motivated by IceCube's observations of shower events with energies between $16$ TeV and a few PeV, which gives the best-fit value of spectrum to be $\Gamma = -2.53 \pm 0.07$~\cite{Aartsen:2020aqd}. The third choice, $\Gamma = -3.0$, is inspired by IceCube's analysis of high-energy starting events, which suggests a spectral index of $\Gamma = -2.89^{+0.20}_{-0.19}$~\cite{Abbasi:2020jmh}.

Utilizing the methodology outlined in the preceding section and expected neutrino events from all considered galaxy mergers calculated using Eqn.~\ref{eq:exp_nu}, we determine TS as a function of $\Phi_0$, the flux normalization of $\nu_\mu+ \bar{\nu}_\mu$ events at 100 TeV, in both uniform and luminosity distance weighting schemes. The outcomes of our stacking analysis are shown in Figs.~\ref{fig:stack_1teV}, \ref{fig:stack_1teV_lum}, and \ref{fig:stack_1teV_w3} for uniform, luminosity distance, and Yuan et al.\,\,weighting schemes, respectively. In the left-hand panel of these figures, we plot TS as a function of $\Phi_0$ for three $\Gamma$ values considered in this work. It is evident from the figures that maximum TS values are near zero in all cases, indicating no significant correlation between galaxy mergers and neutrino events. The cyan shade bands in these plots represent the best-fit values for the astrophysical muon neutrino flux detected by IceCube at $100$ TeV~\cite{IceCube:2021uhz}. The black dashed lines correspond to a $95 \%$ CL upper limit on TS, which provides an upper bound on the muon neutrino flux originating from these galaxy mergers at $100$ TeV. We consider an equal flux assumption of the three neutrino flavors to compute the upper limits on the total neutrino flux at 100 TeV~($\Phi_{3\nu}$), which are presented in Table~\ref{tab:upper_lim}.

\begin{table*}[t]
	\begin{tabular}{ |p{1cm}|p{3.75cm}|p{3.75cm}|p{3.75cm}| }
		\hline
		\multicolumn{4}{|c|}{Upper limit on $\Phi_{3\nu}$ (in units of $\rm GeV^{-1}cm^{-2}sr^{-1}s^{-1}$)} \\
		\hline
	~~~	$\Gamma$ &~~~~ Uniform weighting & ~ Lum. distance weighting & ~~ Yuan \textit{et al.} weighting\\
		\hline
		~~ -2 &~~~~ $ \Phi_{3\nu} \leq 1.11 \times 10^{-18} $&~~~~ $\Phi_{3\nu} \leq 3.69 \times 10^{-19}$  &~~~~$\Phi_{3\nu} \leq 1.02 \times 10^{-18}$\\
	~	-2.53 &~~~~  $ \Phi_{3\nu} \leq{4.44 \times 10^{-19}}$ &~~~~ $ \Phi_{3\nu} \leq 1.44 \times 10^{-19}$ &~~~~$\Phi_{3\nu} \leq 4.01 \times 10^{-19}$\\
	~~ 	-3 &~~~~ $ \Phi_{3\nu} \leq 7.87  \times 10^{-19}$ & ~~~~  $ \Phi_{3\nu} \leq 2.73 \times 10^{-20}$&~~~~$\Phi_{3\nu} \leq 7.08  \times 10^{-20}$ \\
		\hline
	\end{tabular}
	\caption{Upper limits on the total neutrino flux at $100$ TeV ($\Phi_{3\nu}$) that can originate from the galaxy mergers under consideration in the stacking analysis are listed for uniform, luminosity distance, and Yuan \textit{et al.} weighting schemes.}
	\label{tab:upper_lim}
\end{table*}

 In the right-hand panels of Figs.~\ref{fig:stack_1teV}, \ref{fig:stack_1teV_lum}, and \ref{fig:stack_1teV_w3}, we depict the upper limits on the total neutrino flux as a function of energy for three $\Gamma$ values considered in this work. Additionally, we also show the observed total astrophysical diffuse flux of high energy neutrinos measured from the muon-neutrino channel and combined electron and tau neutrino channel at IceCube by cyan shaded regions~\cite{IceCube:2021uhz} and with black data points along with error bars~\cite{Aartsen:2020aqd} respectively. We have also shown the total neutrino flux measured using the PeV energy partially contained events (PEPEs)~\cite{IceCube:2021rpz} by magenta color. The orange-colored bands in these figures represent the latest best-fit astrophysical diffuse neutrino flux obtained using starting track events in IceCube~\cite{Abbasi:2024jro}. It is evident from these figures that the observed flux lies significantly above the 95$\%$ CL upper limits on the flux originating from galaxy mergers examined in this work. We obtain the upper limits of total neutrino fluxes at $100$ TeV using uniform, luminosity distance, and Yuan et al. weighting schemes. For uniform weighting, with $\Gamma=-2$,  the upper limits account for 25.77\%, 22.36\% and 22.09\% of the total astrophysical diffuse flux observed by IceCube measured from muon-neutrino events, combined electron and tau neutrino cascade channels, and starting track events, respectively. Whereas for luminosity distance weighting, with $\Gamma=-2$, the upper limits can contribute no more than 8.55\%,  7.42\% and 7.33\% of the total astrophysical diffuse flux observed by IceCube measured from muon-neutrino events, combined electron and tau neutrino cascade channels, and starting track events, respectively. It is also worth noting that the upper limits on the neutrino flux in the luminosity distance weighting scheme are more stringent than those obtained under the uniform weighting scheme. The majority of mergers are sourced from the catalog J/ApJS/181/233~\cite{2009ApJS..181..233H}, which have $z\sim1$, by far most mergers among all catalogs. Due to the luminosity weighting, the flux is considerably diminished by the $\rm 1/d_L^2$ factor, thus resulting in a significantly lowered value in the measured flux. In the Yuan \textit{et al.} weighting scheme, the upper limits on the flux show that neutrino flux originating from these galaxy mergers can contribute no more than 23.72 \%, 20.57 \% and 20.33\% of the total astrophysical diffuse flux observed by IceCube measured from muon-neutrino events, combined electron and tau neutrino cascade channels, and starting track events, respectively. In this case, the upper limits are not much different as obtained in the uniform weighting scheme. As most of the sources lies around $z\sim1$, they all contribute uniformly to the total flux. While remaining source have $z<1$, which implies that their weight to the total flux will be small as compared to the sources around $z\sim1$. As a result, the limits in Yuan et al.\,\,weighting scheme is slightly lower than those in the uniform weighting scheme.

\section{Conclusions}\label{sec:conc}
The origin of the diffuse high-energy astrophysical neutrino flux observed by IceCube remains a mysterious question. Over many years, several works have been done to address this question. However, we have no definitive, conclusive, and complete answer to this question. This motivates our work, which tests the idea that galaxy mergers can be a potential source of these neutrinos.

For the first time, we explore the idea of galaxy mergers as the possible sources for these high-energy neutrino events. Galaxy mergers can accelerate particles to very high energies through shock mechanism, producing high-energy neutrinos. Due to the weakly interacting nature of neutrinos, we expect them to reach the Earth, and therefore, they can be detected in IceCube. We utilize the galaxy merger data from six catalogs compiled using various observational surveys' results, and 10-year data of muon-track events detected by IceCube to perform a source search analysis of these IceCube observed events.

We perform the search analysis by utilizing the unbinned maximum likelihood ratio method, which has been proven to be a robust tool in such analysis and has been used extensively in the literature. We perform the single and stacking source analyses. In the case of single source analysis, we do not find any significant correlation between galaxy mergers under consideration in this work and high-energy astrophysical neutrino events detected by IceCube. Additionally, a few theoretically motivated galaxy merger sources were also analyzed, but none of them were found to have a significant correlation with the high-energy neutrino events observed by IceCube. For the stacking analysis, we found that the upper bounds of the high-energy neutrino flux for all flavors at $95\%$ CL are $1.11\times 10^{-18}$, $3.69 \times 10^{-19}$, and $1.02 \times 10^{-18}$ $\rm GeV^{-1}\,cm^{-2}\, s^{-1}\,sr^{-1}$ at $100$ TeV with spectral index $\Gamma=-2$. This finding suggests that, while these galaxy mergers may produce high-energy neutrinos, their contribution to the IceCube-detected flux is not substantial.

Our rigorous and robust analysis has led us to conclude that galaxy mergers considered in this work do not predominantly contribute to the high-energy astrophysical neutrino flux detected by IceCube. Hence, the enigmatic nature of these high-energy neutrinos persists, and our quest to find their ultimate sources will continue. We hope that more data on galaxy mergers can shed more light on the question of galaxy mergers being the possible sources of high-energy neutrinos. Moreover, IceCube-Gen-2 and KM3NeT can observe the high-energy neutrino event with more accuracy in their reconstruction direction, which may help us identify the ultimate source of these high-energy neutrinos.

\paragraph*{Acknowledgements\,:}We thank Prerana Biswas, Sovan Chakraborty, Basudeb Dasgupta, Moon Moon Devi, Amol Dighe, Nayantara Gupta, Jagdish C. Joshi,  Albrecht Karle, Joachim Kopp, Reetanjali Moharana, Kohta Murase, Nirmal Raj,  Prantik Sarmah, Prateek Sharma, Pugazhendhi A D, Tracy Slatyer and Bei Zhou for comments and discussions. We especially thank Bei Zhou for detailed discussion regarding our work. S.B. thanks the Council of Scientific and Industrial Research (CSIR), Government of India, for supporting his research under the CSIR Junior/Senior Research Fellowship program through Grant no. 09/0079(15488)/2022-EMR-I. P.P. acknowledges the IOE-IISc fellowship program for financial assistance. M.D. acknowledges the support of the Science and Engineering Research Board (SERB) CRG Grant CRG/2022/004531 for this research. R.L. acknowledges financial support from the Infosys foundation (Bangalore), institute start-up funds, Department of Science and Technology (Govt. of India) for the Grant SRG/2022/001125, and ISRO-IISc STC for the Grant no. ISTC/PHY/RL/499. 
\paragraph*{Software\,:} Python~\cite{10.5555/1593511}, NumPy~\cite{harris2020array}, SciPy~\cite{2020SciPy-NMeth}, Matplotlib~\cite{Hunter:2007}, Astropy~\cite{price2018astropy}, PyAstronomy~\cite{pya}, multiprocessing~\cite{2012arXiv1202.1056M}, iMinuit~\cite{iminuit,James:1975dr}

\bibliography{ref.bib}

\end{document}